\documentclass[hidelinks]{article}

\usepackage{arxiv}

\usepackage[utf8]{inputenc} 
\usepackage[T1]{fontenc}    
\usepackage{hyperref}       
\usepackage{url}            
\usepackage{booktabs}       
\usepackage{amsfonts}       
\usepackage{nicefrac}       
\usepackage{microtype}      
\usepackage{booktabs}
\usepackage{adjustbox}
\usepackage{xcolor,colortbl}
\usepackage{tikz}
\usepackage{textcomp}
\usepackage{hyperref}
\usepackage{lipsum}
\newcommand\copyrighttext{%
  \footnotesize \textcopyright 2020 IEEE. Personal use of this material is permitted.
  Permission from IEEE must be obtained for all other uses, in any current or future 
  media, including reprinting/republishing this material for advertising or promotional 
  purposes, creating new collective works, for resale or redistribution to servers or 
  lists, or reuse of any copyrighted component of this work in other works. 
  DOI: \href{<http://tex.stackexchange.com>}{10.1109/DLS51937.2020.00008}}
\newcommand\copyrightnotice{%
\begin{tikzpicture}[remember picture,overlay]
\node[anchor=south,yshift=10pt] at (current page.south) {\fbox{\parbox{\dimexpr\textwidth-\fboxsep-\fboxrule\relax}{\copyrighttext}}};
\end{tikzpicture}%
}

\definecolor{cadetgrey}{rgb}{0.57, 0.64, 0.69}

\title{Towards a Scalable and Distributed Infrastructure for Deep Learning Applications}

\author{
  Bita Hasheminezhad, Shahrzad Shirzad, Nanmiao Wu, Patrick Diehl\orcid{0000-0003-3922-8419}, Hartmut Kaiser\orcid{0000-0002-8712-2806}\thanks{The STE$||$AR Group - https://stellar-group.org} \\
  Center of Computation \& Technology\\
  Louisiana State University, 
  Baton Rouge, LA, U.S.A.\\
  $\{$bhashe1,sshirz1,wnanmi1$\}$@lsu.edu \\
    $\{$pdiehl,hkaiser$\}$@cct.lsu.edu
   \And
  Hannes Schulz\\
  Microsoft Research Montr\'eal\\
  Montr\'eal, QC, Canada\\
  \texttt{hannes.schulz@microsoft.com} 
}

\newcommand{\tr}{\textsuperscript{\tiny\textregistered}}
\newcommand{\td}{\textsuperscript{\tiny\texttrademark}}

\newcommand{\orcid}[1]{\href{https://orcid.org/#1}{\includegraphics[height=10pt]{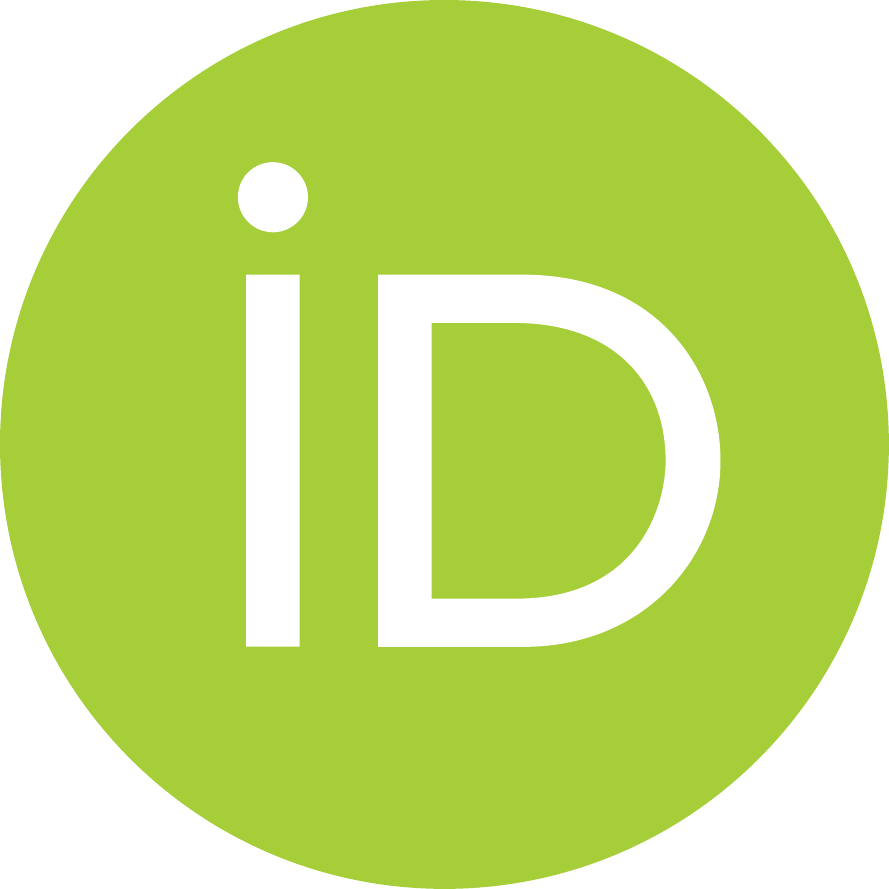}}}

\begin{document}
\copyrightnotice
\maketitle

\begin{abstract}
Although recent scaling up approaches to training deep neural networks have proven to be effective, the computational intensity of large and complex models, as well as the availability of large-scale datasets, require deep learning frameworks to utilize scaling out techniques. Parallelization approaches and distribution requirements are not considered in the preliminary designs of most available distributed deep learning frameworks, and most of them still are not able to perform effective and efficient fine-grained inter-node communication. We present Phylanx that has the potential to alleviate these shortcomings. Phylanx offers a productivity-oriented frontend where user Python code is translated to a futurized execution tree that can be executed efficiently on multiple nodes using the C++ standard library for parallelism and concurrency (HPX), leveraging fine-grained threading and an active messaging task-based runtime system.
\end{abstract}


\section{Introduction}
The recent availability of large-scale datasets and ample computational power has triggered advances in a wide range of Deep Learning (DL) applications. Training a Deep Neural Network (DNN) is a time-consuming iterative process. The slope of advances decreases unless a reasonable training time is maintained. As such, a DL framework must be capable of training the DNN on multiple nodes (be distributed) and efficiently utilize resources on each node.

Most existing DL frameworks are not primarily designed to support major parallelization approaches or work in distributed. In many cases, training a DNN is not investigated as a High Performance Computing (HPC) problem; small models that fit in a single node are developed then scaled as needed. Thus, considerable efforts are required to make those DL frameworks compatible with the requirements of efficient scaling out, namely a distributed address space.

Using HPC capabilities to train DNNs is exceptionally advantageous. To efficiently utilize resources, the DL framework must support overlapping communication and computation. There are HPC frameworks that are designed to overlap communication with proper computation adopting fine-grained parallelism. Message-driven communication, adaptive grain size optimization, and moving work to data are other HPC techniques that can be employed to improve the performance in the DL arena~\cite{awan2017s, peng2019generic, roy2018numa}.

In this paper, we make the following contributions: \textit{1)} we revisit the requirements for scaling the training of DNNs (section \ref{requirements}), \textit{2)} we evaluate the existing distributed DL frameworks based on these requirements (Section \ref{existing}), and finally \textit{3)} we introduce Phylanx as a distributed framework for DL applications. Finally, and we discuss some of Phylanx's design aspects and show primary Scaling Results (Section \ref{phylanx}).

\section{Requirements for Scalable Training of DNNs} \label{requirements}

There is a consensus that DL frameworks must strive to excel at two fundamental classes of characteristics: scalability and ease of use. Many frameworks attempt to put performance and usability at the center of their design~\cite{paszke2019pytorch, sergeev2018horovod}. Some suggest that fine-grained parallelism is the key to achieve high performance~\cite{dryden2019channel, jia2018beyond}. In this section, we elaborate on the requirements for such a DL framework.
\subsection{Scale up and Scale out}
To acquire a reasonable execution time for training modern models on large-scale datasets, it is necessary to exploit every chance of scaling up (scaling on a single node) and scaling out (scaling on multiple nodes) the training process. A single node, even the most powerful one, can never satisfy the memory requirements of contemporary DNNs. In the following, we go through standard parallelization techniques usually applied to divide work across workers (compute resources). After that, we lay out the principles that enable as much resource utilization as possible to improve the parallel efficiency of the training process and decrease the overall time required to train a DNN.

\subsubsection*{\textbf{Deep Learning Parallelization Approaches}}
There are three major approaches for DNN parallelization; Data Parallelism, Model Parallelism, and Pipelining. Hybrid Parallelism is any combination of the aforementioned approaches.

Data Parallelism maintains an identical copy of the entire DNN model in addition to a partition of samples within a minibatch of data on each worker. DNN training is comprised of 3 stages: the \emph{forward pass}, to produce activations and loss using current parameters and labels for the given samples, the \emph{backward pass}, to compute gradients using the loss and parameters generated by the forward pass, and the \emph{solver}, to determine the update rules to update the parameters. In the data parallelism approach, every computational resource independently computes the first two stages. The solver uses a collective communication operation to update the network's parameters at the end of each minibatch of data to keep model replicas consistent. Data parallelism is the most prevalent approach among DL frameworks to scale out the DNN training.

Model Parallelism splits the DNN along an iteration into disjoint or overlapped subsets and trains each subset on one worker. Model parallelism does not have a unanimous definition among the DL community; some define it as splitting the DNN along model parameters wherein a computational resource has only a portion of model parameters~\cite{jia2018beyond}, some specify it as splitting channels or spatial dimensions~\cite{ben2019demystifying}. This paper defines model parallelism as intra-iteration parallelism on any of the tensor dimensions except for the sample dimension. With this broad definition, spatial parallelism is a subset of model parallelism even though it requires all parameters on every computational resource. Model parallelism might incur extra communication. For example, disjoint spatial parallelism requires halo exchange in the forward and backward passes~\cite{dryden2019improving}. Model parallelism is crucial in training novel, wide and deep models. As effectively splitting the DNNs is non-trivial, model parallelism might impose considerable communication overhead if not appropriately designed.

Pipelining is a cross-iteration parallelism approach that assigns each consecutive set of layers in the model to a worker. In very deep neural networks, pipelining the work between workers might be beneficial. The data transfer between workers is comprised of activations and gradients. Pipelining is susceptible to the under-utilization of resources. In a na\"ive implementation of pipelining, only one worker is active at a time due to the sequential dependency of DNNs. The currently active worker performs one of the computationally intensive stages of training on a minibatch, \emph{i.e.}\ forward or backward pass. The next worker has to wait to get the calculated activations or gradients from the current worker. Interleaving microbatch computations, and memorizing the result of calculations were proposed to parallelize this process~\cite{huang2019gpipe, harlap2018pipedream}. 

A pure data parallelization approach is not sufficient for DNN scaling. It is limited by minibatch size and model memory consumption. Although a larger minibatch size means a larger ratio of computation to communication, the minibatch size cannot grow arbitrarily. Memory constraints and accuracy degradation, which slows down convergence and diminishes generalization,~\cite{goyal2017accurate, keskar2016large} impede this growth. Besides, the entire model should fit in the memory even for training on a single sample, which is impossible to achieve for large models. A large-scale DNN training requires hybrid parallelism to scale. This hybrid approach can be a combination of different approaches for different layers of the network, \emph{e.g.}\ data parallelism for layers with few parameters (mainly convolutional layers) and model parallelism for layers that are dense in parameters (fully-connected layers)~\cite{krizhevsky2014one}. A combination of different approaches on each of the layers of the network could be desired to satisfy the necessary granularity~\cite{dryden2019improving, awan2019hypar}. Finally, a hybrid approach can use a distributed storage for model parameters, activations, and gradients at the expense of additional communication~\cite{rasley2020deepspeed}.

\subsubsection*{\textbf{Optimal Resource Utilization}} \label{utilization}
The scalability of a system can be viewed as its ability to efficiently utilize an increasing amount of resources while still satisfying its functionality requirements. In order to scale, a distributed implementation of a DNN training must support overlapping communication and computation, and still maintain the desired training accuracy. Although training a DNN on large datasets is a computation/communication intensive process similar to HPC applications~\cite{ben2019demystifying}, many implementations of parallel paradigms, especially data parallelization, in current DL frameworks have adopted inefficient global barriers imposing unnecessary overheads. Additionally, most existing frameworks are not optimized for CPU clusters, notwithstanding the fact that communication optimization and system resiliency among these are well-established. In this subsection, we present three specifications for a system with optimal resource utilization. The desired framework must use asynchronous collectives, a fine-grained execution platform, and its communication model must be integrated into its internal DL component implementation.

The desired scalable DL framework must utilize asynchronous collectives instead of an asynchronous solver. Updates with Asynchronous Stochastic Gradient Descent (ASGD) were popularized by DistBelief~\cite{dean2012large} and its successor, TensorFlow~\cite{abadi2016tf} to mitigate the straggler effect, meaning computational resources must wait for the slowest machine to complete a phase of computation. The ASGD solver has a low statistical efficiency because parameters in each parameter server are updated by every worker of that server whenever the gradient is ready. As such, one worker might complete an iteration when other workers are updating for the next iteration. To avoid the staleness of gradient updates, ASGD imposes a small learning rate that causes slow convergence and scaling issues~\cite{zhou2017convergence,chen2016revisiting}. That could be the reason that TensorFlow only supports synchronous updates in its distributed data parallelism strategy, MultiWorkerMirroredStrategy\footnote{\url{https://www.tensorflow.org/guide/distributed_training\#multiworkermirroedstrategy}} and its distributed pipelining strategy, GPipe~\cite{huang2019gpipe}. Conversely, synchronous updates with asynchronous collectives would allow the workers to fill the idle-time with useful work.

There are a few conditions for hiding the latencies by using asynchronous collectives: fine-grained but not too small units of work, short context switching times, and significantly reduced synchronization overheads. It is common knowledge in the DL community that a fine-grained execution flow has a higher chance of better scalability. Most DL frameworks support partitioning tensors in the batch dimension; some also support partitioning in other dimensions or a combination of operation dimensions. The desired scalable framework must support partitioning beyond the operation dimensions to be able to utilize the resources efficiently~\cite{jiang2020exploiting}. In practice, tensor partitioning has non-zero overhead and cannot be overused. Also, there is a high variation in gradient sizes, and some of them are too small, far from the scope for optimal communication~\cite{siu2018memory}. Thus, the desired framework, when possible, must combine the small tensors before performing the collective operation~\cite{bao2020preemptive}. Adopting Asynchronous Many-Task (AMT) runtime systems~\cite{kulkarni2019comparative} is a plausible solution to provide fine-grained parallelism beyond hybrid parallelization. Potential AMTs providing distributed computing capabilities are Uintah~\cite{germain2000uintah}, Chapel~\cite{chamberlain2007parallel}, Charm++~\cite{kale1993charm}, Kokkos~\cite{edwards2014kokkos}, Legion~\cite{bauer2012legion}, PaRSEC~\cite{bosilca2013parsec}, and HPX~\cite{Kaiser2020} which are compared in~\cite{thoman2018taxonomy}.

To exploit every opportunity for optimization, the desired scalable DNN framework must be a unified system with communication integrated into DL components' internal implementations. The desired distributed DL framework must contain the functionality to efficiently read the data, preprocess it, and create the DNN model to train its parameters while employing communication between the nodes. It is highly inefficient to run the parts of this workload separately. For instance, ByteScheduler\cite{peng2019generic} is a generic communication scheduler that relies on a dependency proxy mechanism with a non-zero overhead to interact with the DL frameworks execution engines. ByteScheduler improves the performance by partitioning and rearranging tensor transmissions. Still, as it cannot modify the source code of the DL framework that it works on, it must use the DL framework Directed Acyclic Graph (DAG) to create a more refined DAG using different kinds of proxies. HyPar-Flow~\cite{awan2019hypar} is another example of a non-unified system discussed in section~\ref{hypar}.

\subsection{Easy to Use API} \label{easy}
A desired scalable DL platform should have a simple interface that highly abstracts the DL components while being easy to debug. Python has become the de-facto language for the ML/DL community as open-source libraries like NumPy~\cite{walt2011numpy}, SciPy~\cite{virtanen2020scipy}, Pandas~\cite{mckinney2011pandas}, and MatPlotLib~\cite{hunter2007matplotlib} keep up with the high performance numerical analysis and visualization demand. Most DL frameworks have an available API in Python as it is coherent, simple, and readable. Not all of these DL frameworks succeed in presenting a highly abstract or easy-to-debug API, \emph{e.g.}\ newer versions of TensorFlow have adopted Keras~\cite{chollet2015keras} as their default interface. Since debuggability is essential for novel models, the DL community is more receptive to frameworks with imperative paradigms than declarative ones. As such, TensorFlow switched to eager execution as of v2.0.

Besides being intuitive and debuggable, a desired scalable DL framework should have a non-intrusive transition to utilize its parallel and distributed features without requiring the user to add architecture- or platform-specific code or configuration options. A user should not need to modify the DNN model to run it with or without accelerators and should not have to add extra code for setting the parameter servers. Still, the user must decide how many nodes and/or cores they want to use in data and/or model parallelization and/or pipelining, but they need not be aware of the cluster topology to train their DNNs. Therefore, being architecture- and platform-agnostic is an essential characteristic of an easy to use API. Runtime-adaptive capabilities that enable automatic optimizations at runtime are important, such as finding optimal grain-sizes during parallelization or finding the best-possible data distributions depending on the evaluated expressions.

\section{Current Distributed Deep Learning Frameworks} \label{existing}

In this section, we scrutinize existing distributed DL frameworks. We focus on frameworks that are more recent and popular among users. We did not include pure communication schedulers~\cite{peng2019generic, bao2020preemptive,hashemi2018tictac} or frameworks that rely on new hardware~\cite{heydarigorji2020stannis, heydarigorji2020hypertune}. We evaluate these DL frameworks based on the requirements described in Section \ref{requirements}. We summarize our observations in Table~\ref{tab:summary_table} where the columns represent: Data Parallelism, Model Parallelism/Pipelining, Communication overlaps Computation, Sufficient Granularity, Unified, Easy to Use, Architecture Agnostic, License, and Reference. Note that \emph{Sufficient Granularity} represents whether the DL framework can utilize sufficiently fine-grained parallelism. To adequately control the amount of work for the computations (grain size), the DL framework may utilize a combination of parallelization approaches or may use novel solutions. We believe parallelism can and should be applied to the granularity of an individual operation, not just to whole layers~\cite{jia2018beyond}.

\subsection{TensorFlow} 
Although TensorFlow~\cite{abadi2016tensorflow} has always natively supported distributed training, data parallelism, and model partitioning (as of v0.8), older versions of TensorFlow require the user to determine the placement of each operation on devices to run on multiple nodes. Communications scheduling is not supported in TensorFlow out-of-the-box. TensorFlow has a centralized architecture wherein the number of workers cannot grow arbitrarily~\cite{han2019quantitative}. 

TensorFlow also has extensions to support different parallelization approaches. As of v2.2, the Multi Worker Mirrored Strategy is introduced and integrated into TensorFlow for data parallelism. Its update rule is synchronous and it has communication and computation overlapped. Google, the developer of TensorFlow, has developed Mesh TensorFlow~\cite{shazeer2018mesh} and GPipe~\cite{huang2019gpipe} on top of TensorFlow to support model parallelism and pipelining. Many other frameworks are built on top of TensorFlow. Here, we introduce HyPar-Flow, an implementation of hybrid parallelism for DNN training with a Keras-like interface.

\subsubsection{Mesh TensorFlow}
Mesh TensorFlow (MTF) is a language to lay down a general class of tensor computations that requires the cluster of processors to be identical and reliable to perform optimally~\cite{shazeer2018mesh}. Specifying the layout manually, parallelization is possible on any of the tensor dimensions. MTF considers data parallelism as splitting on the batch dimension and enables the user to experiment with any intra-iteration parallelism. A graph of MTF compiles into a SPMD program that depends on MPI communications techniques such as all\_reduce. These all\_reduce-like operations introduce high communication overheads~\cite{huang2019gpipe}.

Utilizing MTF is not straightforward. The user must create the layout and take care of having a reasonable chunk size for each tensor based on the cluster topology (mesh of processors). Also, this layout design is restricted by MTF splitting rules, \emph{e.g.}\ two dimensions of a tensor are not allowed to be split across the same mesh dimension. Some available features of MTF are only tested on accelerators and, in particular, Tensor Processing Units (TPUs). In some cases, although the GPUs are detected, most computations are run on CPUs. MTF is not compatible with the newer versions of TensorFlow (eager evaluations).

\subsubsection{GPipe}
GPipe is a pipeline parallelism library implemented under Lingvo~\cite{shen2019lingvo} framework which is a TensorFlow framework focusing on sequence-to-sequence models. GPipe partitions operation in the forward and backward pass and allows data transfer between neighboring partitions. Therefore, there is a lower bubble overhead in comparison to a naive pipeline parallelization. GPipe does not memorize the activations of layers. As such, it can scale for large models as the number of accelerators may increase immensely, however, it needs to re-compute the activations of layers for the backward pass. In other words, GPipe is a scalable solution that attempts to have a higher utilization in a pipeline, but a portion of this utilization belongs to the re-computation. GPipe can be combined with data parallelism for a better scale.

There are few examples of GPipe available. It is unknown if GPipe is compatible with the eager TensorFlow. GPipe works only on accelerators and it assumes a single layer fits in the memory of one accelerator.

\subsubsection{HyPar-Flow} \label{hypar} HyPar-Flow is an implementation of data, model, and hybrid (combination of data and model) parallelization on Eager TensorFlow using MPI for communication and Keras for interface\cite{awan2019hypar}. HyPar-Flow uses Horovod (Section\ref{horovod}) for pure data parallelism. It implements the hybrid parallelism by generating a representation of the Keras code in a distributed manner and an MPI communicator for each model partition to devise the communication and computation overlap. HyPar-Flow is a non-unified framework, and it recognizes TensorFlow as a separate unmodifiable framework. Since in TensorFlow there is no access to the gradients of other layers, HyPar-Flow has to add a layer-like structure before each TensorFlow's layer.

HyPar-Flow only requires the strategy, the number of model partitions, and the number of model replicas from the user to utilize them with every possible intra-iteration parallelization. Its debuggability depends on its backend, TensorFlow. HyPar-Flow is platform-agnostic and also optimized for many-core CPU clusters, too.

\subsection{Caffe} Berkeley AI Research founded Convolutional Architecture for Fast Feature Embedding (Caffe)~\cite{jia2014caffe} DL framework which does not support distributed training out-of-the-box. Caffe is a define-and-run (declarative) framework that has three blocking phases for training a DNN. There are many extensions of Caffe that support distributed training centralized or decentralized; each focuses on a specific platform or communication library. FireCaffe~\cite{iandola2016firecaffe} and MPI-Caffe~\cite{lee2015m} use MPI to respectively deploy data and model parallelism on multi-GPU clusters. FeCaffe~\cite{he2019fecaffe} and Caffe Barista~\cite{vink2020caffe} are FPGA specializations of Caffe for Convolutional Neural Networks (CNNs). Intel-Caffe\footnote{\url{https://github.com/intel/caffe}} supports data parallelism training on CPU-based clusters. NVIDIA\tr-Caffe\footnote{\url{https://ngc.nvidia.com/catalog/containers/nvidia:caffe}} is not a distributed framework. Here we discuss S-Caffe~\cite{awan2017s} and NUMA-Caffe~\cite{roy2018numa} further.

\subsubsection{S-Caffe}
S-Caffe or OSU-Caffe is a product of co-designing CUDA-Aware MPI runtime and Caffe for data parallelism on GPU clusters. To overlap three phases of training in Caffe, S-Caffe uses on-demand communication and multi-stage non-blocking collectives. Except for a helper thread in gradient aggregation, S-Caffe does not use multi-threading.

\subsubsection{NUMA-Caffe} NUMA-Caffe is a distributed DL framework that compliments BVLC-Caffe and Intel-Caffe implementations. BLVC-Caffe and Intel-Caffe apply BLAS-level and batch-level parallelism, respectively. NUMA-Caffe adds two additional levels of parallelism: Non-Uniform Memory Access (NUMA) node-level parallelism and thread-level parallelism. It also eliminates many remote memory accesses providing automatic data localization. NUMA-Caffe is platform-specific and it works around Caffe's implementation and its focus is Convolutional Neural Networks (CNNs).

\subsection{PyTorch DDP} PyTorch, a successor of Caffe2\footnote{\url{https://caffe2.ai}}\label{sec:pytorchDDP}, is an imperative DL framework developed by Facebook using dynamic computation graphs and automatic differentiation~\cite{paszke2019pytorch}. PyTorch is easy to use, debug, and customize. However, it functions such that $10$\% of speed can be traded in order to acquire a considerably intuitive model. Some PyTorch designs are admittedly susceptible to certain corner cases. PyTorch Distributed Data Parallel~\cite{li2020pytorch} is an extra feature intercepted to PyTorch to perform DDP computations and is available as of v1.5. PyTorch RPC is developed to support model parallelism but, as of now, is a project in progress.

PyTorch DDP utilizes some techniques that are engineered to increase performance based on practice. These techniques are gradient bucketing (adds a hyper-parameter, bucket, to launch each all\_reduce. Small tensors bucket into one all\_reduce operation), overlapping communication with computation (which depends on when the first bucket gets ready and the backward computation order), and skipping synchronization. As these techniques are tuned to practice, PyTorch admittedly asserts that some of the solutions are not perfect but are reliable approximations on with minimum overhead. These techniques and shortcomings are discussed in~\cite{li2020pytorch}. PyTorch mainly focuses on ease of use, and enables users with options in training their models. For instance, for trainings that require larger scales, developers can explore enabling no\_sync mode (skipping synchronization) if it attains acceptable convergence speed.

\subsection{Horovod} \label{horovod}
Horovod~\cite{sergeev2018horovod} is a stand-alone Python library for data parallelism using an optimized ring\_allreduce collective and a tensor fusion algorithm that works on top of another DL framework. At first, Horovod only supported TensorFlow as the DL worker, but currently, it supports PyTorch and MXNET, too. Horovod completely replaces the parameter server-based optimizer of TensorFlow which underutilizes the resources because of its communication overhead~\cite{awan2018scalable} with its synchronous optimizer. It can almost achieve linear performance gain if the portion of parameters in the dense layers to all parameters is small, \emph{e.g.}\, unlike VGG-16~\cite{simonyan2014very}, the dense layer of ResNet50~\cite{he2016deep} has a small portion of all parameters. Horovod supports model partitioning but does not support model or pipeline parallelism, so it can train only models that fit into a single device (maybe with multiple GPUs). Although it has one of the most optimized asynchronous collectives, in the absence of granularity, the communication overhead significantly grows with the number of nodes~\cite{wu2018performance}.

Horovod has a simple API and its own profiling tool, Horovod Timeline. The transition to distributed on TensorFlow code is easier than the original TensorFlow code transition. It also has a Keras interface which is popular in the DL community. Horovod is optimized for GPUs but can work without GPUs as well.

\subsection{FlexFlow} FlexFlow is a DL framework with an execution optimizer that can find a strategy for any intra-iteration hybrid parallelization~\cite{jia2018beyond}. The execution simulator is initialized with data parallelism as well as anther parallelization strategy that is selected at random. Unless it takes more than the time budget of searching, it can find the optimal strategy. While this strategy is the best for intra-iteration parallelization, communication between partitions is not necessarily optimal. FlexFlow is built on an AMT, the Legion~\cite{bauer2012legion} runtime and it is able to parallelize its task graph at the granularity of a single operation. 

FlexFlow has a Keras-like and a native interface. Using FlexFlow's execution optimizer is rather simple; it only takes the cluster topology and the graph corresponding to the DNN. FlexFlow works only on GPUs, though currently many top clusters in the world are not equipped. 

\subsection{Chainer} 
Chainer is the pioneer DL framework with a Define-by-Run (imperative) paradigm~\cite{tokui2019chainer}. PyTorch is inspired by Chainer but focuses more on performance; PyTorch DDP passes most of the intensive computations to its C++ core while Chainer is developed purely in Python. Chainer only supports data parallelism. It has a synchronous decentralized design that can be realized by all\_reduce communication while communication scheduling is not supported.

Chainer is simple to use and debug, and this simplicity has been the focal point of its implementation. Basic knowledge of Python and neural networks is sufficient to use it. GPU runs are supported through CuPy and allow users to code in a CPU/GPU-agnostic environment~\cite{tokui2015chainer}. ChainerCV is an add-on package specialized for computer vision tasks for prototyping ideas by non-experts.
\subsection{CNTK} 
Computational Network ToolKit (CNTK)~\cite{seide2016cntk} is Microsoft's open-source declarative library for DL. CNTK has developed four algorithms for its solver: Data-Parallel SGD, Block Momentum SGD, Model Averaging SGD, and Data Parallel ASGD\footnote{\url{https://docs.microsoft.com/en-us/cognitive-toolkit/multiple-gpus-and-machines}}. Data-Parallel SGD uses a trick for reducing the size of communication called 1-bit SGD which compresses the gradient values to a single bit. The difference between the original gradient and its quantized value is added to the next minibatch. Block Momentum SGD uses Blockwise Model Update and Filtering (BUMF) which requires resetting momentum to zero while maintaining frequent model synchronization to converge. Microsoft\td{} no longer recommends Model Averaging SGD since it falls behind the Data-Parallel SGD and Block Momentum SGD. Data-Parallel ASGD can be useful for models that are less sensitive to noise. CNTK does not support model parallelism.

CNTK has APIs in Python, C\#, C++, and BrainScript which is its domain specific language for defining networks. It provides a performance profiler in Python that generates a detailed profile log. Users can get more information to debug by plotting the underlying graph easily with the logging information\footnote{\url{https://cntk.ai/pythondocs/Manual_How_to_debug.html}}. CNTK has been one of the official Keras' backends as of v2.0. By changing the argument in the device setting, CNTK can easily switch between its CPUs and GPUs implementations.

\subsection{BigDL} BigDL is a distributed DL framework for data parallelism on top of Spark~\cite{dai2019bigdl}. It does not support model parallelism. Like TensorFlow, BigDL has a parameter server-style architecture, but unlike TensorFlow, it favors coarse-grained operations where data transformations are immutable. BigDL processes its calculations across the nodes through the InvokeAndWait function. On a CPU cluster, BigDL is faster in computations than TensorFlow, benefiting from its CPU optimizations, but it suffers from long communication delays due to its dependency on MapReduce framework~\cite{du2018comparative}.

BigDL is integrated into the Spark functional compute model and does not suffer from overheads due to the adaptation of frameworks. It has been developed by Intel\td{} Analytics to handle the stream of dynamic and messy data. TensorFlowOnSpark~\cite{tfOnSpark2017} and CaffeOnSpark~\cite{CaffeOnSpark2017} use a connector approach to connect to Spark, and both have an execution model that is very different from Spark. BigDL runs a series of a-couple-of-seconds Spark jobs which are scheduled by Spark. On large clusters, scheduling may become a bottleneck for Spark; therefore, on each server, BigDL is set to not launch more than one task with multiple threads in each iteration. A benefit of using Spark is that it is equipped with fault tolerance and a fair load balancing mechanism. BigDL is modeled after Torch\footnote{\url{http://torch.ch/}} and is easy to use and debug. It works best on a single Intel\td{} Xeon CPU node.

\subsection{SINGA} 
SINGA~\cite{ooi2015singa} is an imperative distributed DL framework and one of the few that is primarily designed considering scaling. SINGA supports different synchronous (Sandblaster and AllReduce) and asynchronous (Downpour and Distributed Hogwild) solvers and is able to run with different workers/servers configurations. It can have multiple worker/server groups running asynchronously while each group of workers runs synchronously; therefore, SINGA is not limited to the medium-sized clusters. SINGA is not well optimized, it assigns one thread to each processor. SINGA exhibits good scaling out results but lacks in scaling up and accuracy~\cite{shams2017evaluation, wang2015singa}. It supports data and model parallelism.

Users of SINGA should have a clear understanding of its layer-based programming model. They can choose to run their program on multiple nodes, but they need to configure it first. SINGA has several built-in layers. Also, layers are customizable as long as they are consistent with TrainOnBatch algorithm. SINGA is one of the few frameworks which allows the user to manually partition the layer transparently using concatenation and slice layers; it is one of the aspects of its design for scalability. SINGA has APIs in C++ and Python and runs on clusters with or without accelerators.

\subsection{MXNET-MPI}
MXNET-MPI~\cite{mamidala2018mxnet} is the extension of MXNET that replaces each worker in a parameter server architecture with a group of workers. Workers of each group are synced together using an MPI collective operation where the solver is synchronous SGD. Therefore, for data parallelism, MXNET-MPI performs better than a fully centralized parameter server dependent architecture or a fully decentralized architecture that synchronizes parameters in a blocking fashion. Thus, overall execution time is reduced, even though there is no explicit scheduling for overlapping communication and computation. Putting workers into groups and having both synchronous and asynchronous updates is similar to SINGA's design. MXNET does not support multi-node model parallelism. 

The key-value store is the critical component for training the DNN with MXNET on multiple nodes. The v1.5.1 MXNET, does not support training on more than 4 GPUs and, the set-up for distributed training is not user-friendly~\cite{mahon2020performance}.


\subsection{DeepSpeed / ZeRO}
ZeRO~\cite{rajbhandari2019zero} approaches training large models by focusing on solving the memory limitation problem while attempting to minimize the overhead. To train models where the memory of a single GPU is the limiting factor, ZeRO partitions activations, optimizer states, gradients, and parameters and distributes them equally overall available nodes. It then employs overlapping collective operations to reconstruct the tensors as needed. This gives DeepSpeed the memory advantages of model parallelism and pipelining, while retaining the flexibility and ease of use of data parallelism.

The published DeepSpeed~\cite{rasley2020deepspeed} implementation can be used as a drop-in replacement for PyTorch's DDP (Section~\ref{sec:pytorchDDP}) and can optimize any operation that is derived from the \texttt{pytorch.nn.module}. Due to this fine granularity, DeepSpeed can make tailored decisions about which tensors to distribute, whether to off-load memory to CPU and where to limit buffer sizes to prevent out-of-memory issues in the allocator.
While the current release version does not include model parallelism or pipelining techniques, DeepSpeed is compatible with approaches that do.

\begin{table*}[tbp]
    \caption{Comparison between Distributed Deep Learning Frameworks}
    \begin{adjustbox}{width=1\textwidth}
    \centering
    \begin{tabular}{@{}lccccccccc@{}}
    \toprule
    \textbf{Framework} & \textbf{Data Par} & \textbf{Model Par/Pipeline} & \textbf{Comm overlaps Comp} & \textbf{Sufficient Granularity} & \textbf{Unified} & \textbf{Easy to Use} & \textbf{Archt. Agnostic} & \textbf{License} & \textbf{Reference} \\\midrule
    Mesh-TensorFlow & \checkmark & \checkmark & & & & (-) & & Apache-2.0 & \cite{shazeer2018mesh}  \\\hline
    GPipe & compatible & \checkmark & \checkmark & &\checkmark & (-) & & Apache-2.0 & \cite{huang2019gpipe}  \\\hline
    HyPar-Flow & \checkmark & \checkmark & \checkmark & \checkmark & & (++) & \checkmark &  N/A  & \cite{awan2019hypar}  \\\hline
    S-Caffe/OSU-Caffe & \checkmark & &  \checkmark & & \checkmark & N/A &   & N/A$^{*}$ & \cite{awan2017s}  \\\hline
    NUMA-Caffe & \checkmark & &  \checkmark & & & (+) &  & Public Domain & \cite{roy2018numa}  \\\hline
    PyTorch DDP & \checkmark & &  \checkmark & & \checkmark & (++) & \checkmark  & BSD-style & \cite{li2020pytorch}  \\\hline
    Hororvod & \checkmark & compatible &  \checkmark & & & (++) & \checkmark &  Apache-2.0 & \cite{sergeev2018horovod} \\\hline
    FlexFlow & \checkmark &  \checkmark & \checkmark & \checkmark & \checkmark & (+) &  & Apache-2.0 &  \cite{jia2018beyond}  \\\hline
    Chainer & \checkmark &  & & & \checkmark & (++) & \checkmark  & MIT &  \cite{tokui2019chainer}  \\\hline
    CNTK   & \checkmark & & \checkmark & & \checkmark & (++)  & \checkmark & MIT & \cite{seide2016cntk} \\\hline
    BigDL & \checkmark & & & & \checkmark & (+) &  & Apache-2.0 & \cite{dai2019bigdl}  \\\hline
    SINGA & \checkmark & \checkmark & & & \checkmark & (+) & \checkmark &  Apache-2.0 & \cite{ooi2015singa}  \\\hline 
    MXNET-MPI & \checkmark & &  & & \checkmark & (++)& & Apache-2.0 & \cite{mamidala2018mxnet}  \\\hline
    DeepSpeed &  \checkmark & compatible &  \checkmark & \checkmark &  \checkmark & (++)&  &  MIT & \cite{rasley2020deepspeed}  \\\hline
    \rowcolor{cadetgrey}
    Phylanx & \checkmark & \checkmark & \checkmark & \checkmark & \checkmark & (++) & \checkmark  & Boost-1.0 & \cite{8638482}  \label{tab:summary_table}\\\bottomrule
    \multicolumn{2}{l}{$^{*}$Only available in binary form}
    \end{tabular}
    \end{adjustbox}\\[2mm]
    \scriptsize{\textbf{Data Par} shows if the framework supports Data Parallelism on multiple nodes. \textbf{Model Par/Pipeline} is checkmarked if the framework supports any network intra-iteration or inter-iteration parallelism other than data parallelism. \textbf{Comm overlaps Comp} highlights if there is any explicit attempt that prevents sequential run of computation and communication. \textbf{Sufficient Granularity} represents whether parallelism is applicable to the granularity of an individual operation. \textbf{Unified} is checkmarked if the infrastructure that makes the training distributed is integrated into the implementation of its DL components. \textbf{Easy to Use} has two measures: simple interface for coding and debugging. \textbf{Archt. Agnostic} means no modification is needed to use the code in distributed on different architectures or platforms. \textbf{License} and \textbf{Reference} are self-explanatory.}
    
\end{table*}

\section{Phylanx as a Deep Learning Framework} \label{phylanx}
Phylanx~\cite{phylanx_github,8638482} is an asynchronous distributed array processing framework that combines the benefits of exposing a high-productivity Python-based development environment with a high-performance C++-based execution environment targeting computer systems of any size, including cloud and HPC-cluster technologies. The framework automatically transforms the user-provided Python code into an intermediate representation that is efficiently executed and distributed across all available compute resources as specified by the user. 

\subsection{How Phylanx tackles design challenges of DL frameworks}
Phylanx is a software framework that is based on the HPX runtime system, which in turn was designed from first principles to address well known key challenges of high-performance computing applications. As such, Phylanx implicitly and naturally benefits from inheriting the advantages of applying fine-grained parallelism, message driven computation, constrained-based synchronization (never synchronize more than needed for the local progress of execution), implicit overlapping computation with communication, and runtime-adaptive granularity control. In this section we describe these capabilities in more detail and demonstrate that Phylanx addresses some of the identified challenges of DL frameworks (see Table~\ref{tab:summary_table}). We are aware that Phylanx is no ``jack of all trades'' solution, but many of the challenges are tackled.
\subsubsection*{\textbf{DL Paralleization approaches}}
In Phylanx, the distribution of data is done by tiling the data arrays involved. Each locality the application runs on is responsible for one (or more) tiles of the data. Phylanx execution is strictly SPMD-style, \emph{i.e.}\ each locality executes a structurally equivalent expression graph, while each part of that graph performs communication between the localities depending on the operations to be performed. The execution of this expression graph is done fully asynchronously, and the necessary communication relies on asynchronous collectives. This allows to fully overlap them with the ongoing computations. This approach allows to seamlessly distribute the processed data arrays across a possibly large amount of localities (\emph{i.e.}\ nodes in a cluster) while maintaining the best possible scalability. Each of the tiles of the data arrays handled by a locality is internally represented exactly like a fully local data array except that it carries additional meta-information describing the whole (distributed) array. This has the benefit of simplifying the implementation.

Phylanx supports overlapped tiling, which is beneficial in spatial parallelization. A halo exchange is needed in forward and backward pass using spatial parallelization. When the kernel size is comparably smaller than the data size, the user can use overlapped tiling to avoid extra communication.

\subsubsection*{\textbf{Communication overlaps Computation}}
Phylanx uses the HPX runtime system as its underlying execution platform. HPX is the C++ Standard Library for Parallelism and Concurrency~\cite{hartmut_kaiser_2020_3675272,Heller2017,Kaiser2020}. It provides the necessary abstractions to build efficient codes that are oblivious to local and remote operations while maintaining efficient data locality. HPX has been described in detail in other publications~\cite{heller:2012,Heller:2013:UHL:2530268.2530269,hpx_pgas_2014,Kaiser:2015:HPL:2832241.2832244,Heller2016}. In the context of Phylanx, we use HPX because of its dynamic scheduling and global data addressing capabilities as well as its ISO C++ standards conforming API. The constructs it exposes are well-aligned with the existing C++ standards~\cite{cxx11_standard,cxx14_standard,cxx17_standard,cxx20_standard}. HPX is fundamental for Phylanx, as it uses shared memory abstractions that have already been adopted in the most recent ISO C++ standards and HPX's distributed memory abstractions are standards conforming extensions. Using the \emph{Futurization} technique in HPX, the execution of Phylanx code is expressed as complex dataflow execution graphs that can generate a large amount of fine-grained parallel tasks that are scheduled to execute only when their dependencies are satisfied (see for instance~\cite{heller19harnessing}). This naturally and intrinsically enables overlapping computation with communication, thus perfectly hiding communication latencies.

Phylanx achieves overlapping computation and communication using its asynchronous active messaging communication platform (that also includes its asynchronous collectives). Phylanx prefers moving work to data over moving data to work and to means of runtime-adaptively coalesce messages into larger units (tensor fusion)~\cite{10.1145/3337821.3337915,wagle2018methodology}, which further reduces the latencies and overheads caused by the necessary communication operations (see Section~\ref{subsubsec:grainsize} for more information).

\subsubsection*{\textbf{Sufficient Granularity}}
\label{subsubsec:grainsize}
Most of the analyzed DL frameworks have mentioned fine-grained parallelism. The purpose is to have a synchronous optimizer while maintaining acceptable performance. However, with fine-grained parallelism, meaning many small computational tasks are executed, the overhead of context switching takes into account using the system threads. 
 
Phylanx utilizes HPX's lightweight user-level threading system to reduce the context switching and synchronization overheads of small computational tasks to its minimum~\cite{kaiser2009parallex, laberge2019scheduling}. HPX's asynchronous execution model combined with its active-message based communication model intrinsically allows to hide latencies exhibited by those communication operations, naturally overlaps those with proper computation, and reduces synchronization overheads across nodes to a minimum. HPX's asynchronous collectives allow to break the strict lock-stepping between ranks and enable them to perform other work while the collective operation is being performed.

In order to avoid for threads to become too short lived (which increases associated overheads), Phylanx employs runtime-adaptive techniques for controlling the granularity of tasks and the size of networking packages, both with the goal of reducing overheads and maximizing system utilization~\cite{wagle2019runtime}.

\subsubsection*{\textbf{Unified}}
Phylanx exploits every chance of parallelism since it implements every required DL operation using HPX. Thus, the infrastructure that makes the training scale out is integrated into the DL framework. On the contrary, we observe that not considering scaling out requirements in designing a DL framework has left us stitching different libraries (connector approach) or re-implementing components to achieve functionality.

\subsubsection*{\textbf{Easy to Use}}
Phylanx provides a high-productivity debuggable Python-based interactive interface, JetLag\footnote{\url{https://github.com/STEllAR-GROUP/JetLag}} ~\cite{brandt2020jetlag}. JetLag constitutes a container-based Jupyter notebook interface, a performance visualization platform~\cite{williams2019visualizing} (Traveler) and a performance measurement library (APEX~\cite{10.5555/3026759.3026764}). The user can code in a Jupyter notebook and easily plot node-link diagram of the execution tree as well as Gantt and utilization charts using APEX performance counters~\cite{10.1145/3337821.3337915}.
\subsubsection*{\textbf{Architecture Agnostic}}
Running Phylanx on shared memory or distributed memory is hidden from the user due to the high-level abstractions provided by HPX. A unified syntax and semantics for local and remote operations provided by the Adaptive Global Address Space (AGAS)~\cite{amini2019assessing,kaiser2014hpx} is utilized. Thus, the user does not need to modify the Python code for local and remote computations or use another platform.
    
In addition to addressing the above challenges, Phylanx supports fault tolerance. Long-running non-dedicated resources can be used to train a large DNN; therefore, a means of fault tolerance is essential in the occurrence of a failure~\cite{abadi2016tf}. TensorFlow provides a checkpoint-based fault tolerance system while BigDL incorporates fault tolerance using Spark's RDD~\cite{zaharia2012resilient}  on every operation. HPX provides software resilience~\cite{gupta2020implementing} to implement fault tolerance within Phylanx as well. HPX can detect silent data corruptions, \emph{e.g.}\ memory bit flips or CPU floating point errors. After some corrupted computation on a node, the user can do one of the following: \textit{1)} replay the computation and use the new computation if the silent data corruption vanished. \textit{2)} start replicates of the computation which are executed independently. There are three possibilities to compare the replicates: \textit{a)} use check sums to compare; \textit{b)} use a user-defined consensus function, which returns the replicate passing the tests; and \textit{c)} if multiple replicates pass the consensus function, the user provides a validate function to decide which one to use. 

HPX aims to resolve the problems of scalability, resiliency, power efficiency, and runtime adaptive resource management that continue to grow in importance, as the industry is facing increasing demands in supporting highly distributed heterogeneous systems. 

\subsection{Primary Scaling Results}
\begin{figure}
  \centering
  \includegraphics[width=0.5\textwidth]{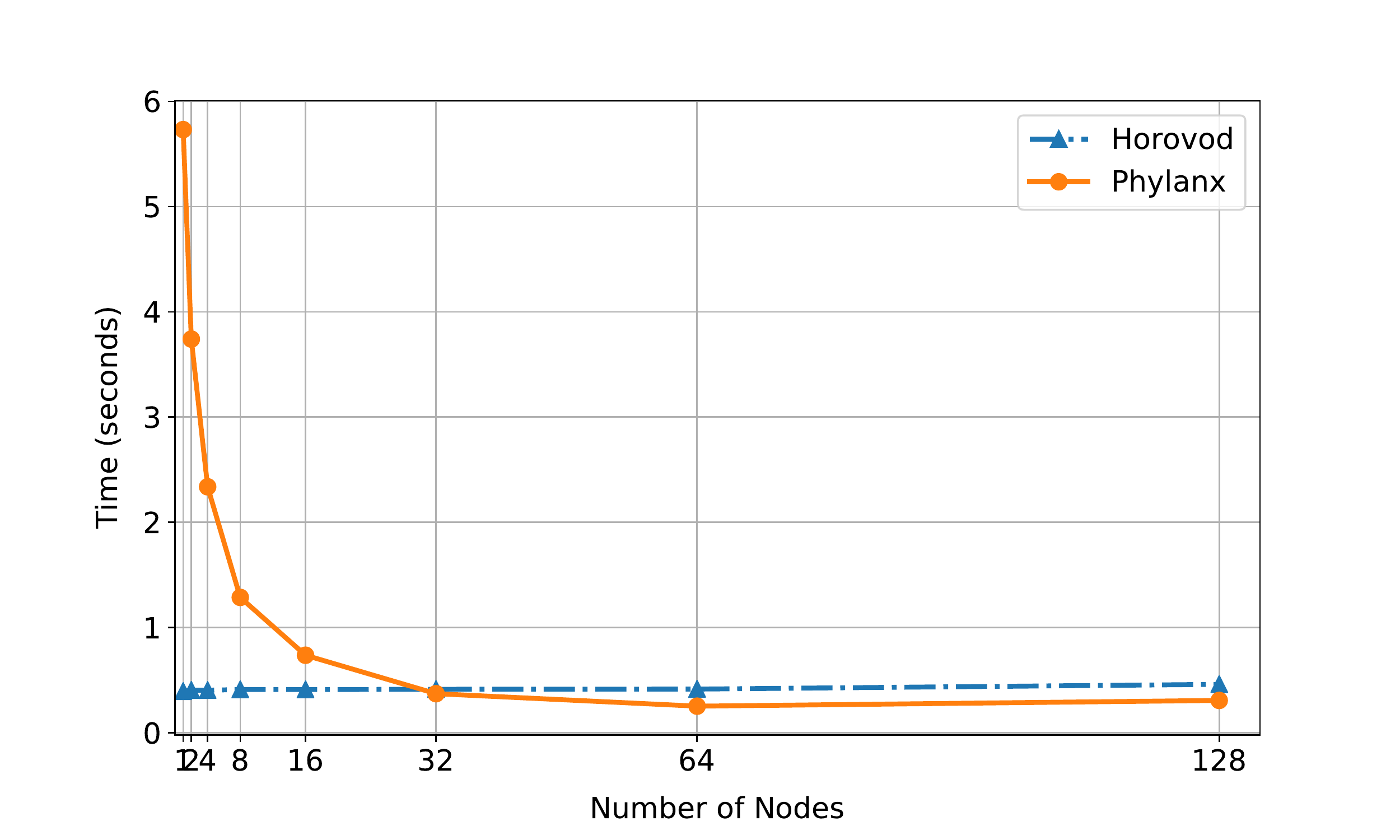}
  \caption{Comparison of Phylanx and Horovod executing forward propagation of a CNN on an HPC cluster up to $128$ nodes.}\label{fig:perf}
\end{figure}

We have used an Intel\td{} Cascade Lake based cluster called Queen Bee $3$ maintained by LONI~\cite{loni}. This CPU cluster consists of $192$ nodes each containing $2{\times}24$-Core Xeon Platinium $8260$ processors. We have measured the execution time for the forward pass of a $4$-layers CNN (deduced from Kaggle\footnote{\url{https://www.kaggle.com/niyati11/project-convo1d}}) on a Human Activity Recognition data. For a mini-batch size of $8000$, we compare the execution time of Phylanx (git hash: 03ad3b8) to Horovod v0.20.0, which is installed on top of TensorFlow v2.3.0 and uses Gloo (git hash: fe2ad9c)~\cite{incubator2017gloo} as its communication library in Figure~\ref{fig:perf}. First, it can be seen that the execution time of Horovod does not significantly decrease when using more nodes while that of Phylanx shows a notable reduction, which demonstrates the scalability of Phylanx. Second, we find that Phylanx takes a shorter execution time at least ($\approx 18$\%) in comparison to Horovod when using 32 or more nodes, which provides insight into running CNN on larger clusters. 

\section{Conclusion}
With the ever-increasing need for reducing the time of training modern deep neural networks, scaling out has become the standard practice. To this effect, we revisited the requirements for distributed training of deep neural networks and described the underlying factors for the desired framework. It must provide \textit{a)} asynchronous collectives, \textit{b)} fine-grained execution platform, \textit{c)} must be unified \textit{d)} must be easy to use and debug, and \textit{e)} must be architecture and platform agnostic. Most existing frameworks started as single-node computational models then adapted to running on GPU clusters. However, we should design a system from first principles to embrace the challenges of platform-agnostic distributed computation. We showed that Phylanx offers great potential even in its current state.

\section*{Acknowledgements}
The authors are grateful for the support of this work by the LSU Center for Computation \& Technology and by the DTIC project: Phylanx Engine Enhancement and Visualizations Development (Contract Number: FA8075-14-D-0002/0007).

\bibliographystyle{unsrt}  
\bibliography{bib}  

\end{document}